\documentclass[intlimits,twoside,a4paper]{article}
\usepackage{amsmath,amssymb}
\usepackage{graphicx}
\usepackage{multirow}
\usepackage[english]{babel}
\usepackage{color}
\usepackage{caption}
\usepackage{subcaption}
\usepackage[usenames,dvipsnames,svgnames,table]{xcolor}
\usepackage[T2A]{fontenc}
\usepackage[cp1251]{inputenc}

\usepackage[eqsecnum]{cmpj2}

\issue{2016}{19}{2}{23601}
\doinumber{10.5488/CMP.19.23601}

\title[Salt-specific effects in lysozyme solutions]
{Salt-specific effects in lysozyme solutions\thanks{We dedicate this contribution to our friend and coworker Professor A.D.J.~Haymet on occasion of his 60$^\text{th}$ birthday.}}
\author[T. Janc \textsl{et al.}]
{T. Janc,  M. Kastelic, M. Bon\v cina, V. Vlachy}
\address{Faculty of Chemistry and Chemical Technology, University of
Ljubljana, Ve\v{c}na pot 113, 1000 Ljubljana, Slovenia}

\date{Received November 12, 2015}
\authorcopyright{T. Janc,  M. Kastelic, M. Bon\v cina, V. Vlachy, 2016}

\begin{document}

\maketitle


\begin{abstract}
The effects of additions of low-molecular-mass salts on the properties of
 aqueous lysozyme solutions are examined by using the
cloud-point temperature, $T_\textrm{cloud}$, measurements.
 Mixtures of protein, buffer, and simple salt in water are
studied at \mbox{$\text{pH}=6.8$} (phosphate buffer) and
\mbox{$\text{pH}=4.6$} (acetate buffer). We show that an
addition of buffer in the amount above \mbox{$I_\textrm{buffer}
= 0.6$ mol\hspace{2pt}dm$^{-3}$}
does not affect the \mbox{$T_\textrm{cloud}$} values. However,
by replacing a certain amount of the buffer electrolyte by
another salt, keeping the total ionic strength constant, we can
 significantly change the cloud-point temperature. All the salts
 de-stabilize the solution and the magnitude of the effect
 depends on the nature of the salt. Experimental results
 are analyzed within the framework of the one-component
 model, which treats the protein-protein interaction as highly
 directional and of short-range. We use this approach to predict the second virial coefficients, and liquid-liquid phase
 diagrams under conditions, where \mbox{$T_\textrm{cloud}$} is
 determined experimentally.
\pacs 64.60.My, 64.70.Ja, 87.15.km, 87.15.nr, 87.80.Dj
\end{abstract}


\section{Introduction}

Studies of the physico-chemical behaviour of mixtures of
proteins and simple salts in
water \cite{Collins1997,Tavares2004a,Finet2004,Zhang2006,Zhang2009,
Kunz2009,Kunz2010,Ninham2010,Parsons2011,Lonostro2012,Medda2013,
Jungwirth2014} are important to understand the stability of
such mixtures and may yield improvements of methods for protein
precipitation and
crystallization \cite{Thomson1987,Kautt1989,Taratuta1990,Kautt1992,
Coen1995,Broide1996,Rosenbaum1996,Lomakin1999,Rosenbaum1999,
Grigsby2001,Curtis1998,Sear1999,Moon2000,Curtis2002,Tavares2004,
Tavares2004b,Tavares2004c,Bostrom2005,Zhang2005,Bianchi2006,Lima2007,
Annunziata2008,Mehta2012,Lezhang2012,Quang2014,Kastelic2015,Gunton2007}.
The list of references presented here is far from being complete~---
there are more papers published than we can possibly mention.
Protein aggregation may be desired or undesired. In the
downstream processing, proteins should be salted-out in such a
way that their native form is preserved.  From the undesired viewpoint:
the bio-pharmaceutical formulations should be free of aggregates,
and their formation must be inhibited during
storage \cite{Frokjaer2005}. Further, (the pathological)
protein aggregation appears to be connected with several
diseases \cite{Chiti2006}. A better understanding of the factors
that influence the aggregation of proteins in a mixture with various
salts is, therefore, of great importance.

The temperature-induced liquid-liquid phase separation in an
aqueous protein-water solution was performed for lysozyme solutions,
first reported by Ishimoto and Tanaka \cite{Ishimoto1977}.
So far the method has been used many
times \cite{Thomson1987,Taratuta1990,Broide1996,Grigsby2001,Zhang2009},
having proved to be useful in studying the salt-specific effects in
protein solutions. Having cooled the protein solution at a
constant concentration, a well-defined and reversible
opacification is observed when a certain temperature, named
$T_\textrm{cloud}$, is reached. Above this temperature, the solution
exists in one phase, while below the $T_\textrm{cloud}$, two
equilibrium phases are observed. The onset of the cloud-point
temperature depends, not only on $\text{pH}$ of the solution,
but also on the electrolyte concentration and its nature or
composition if several salts are present. For this reason, the
$T_\textrm{cloud}$ measurements represent a useful tool in the studies of
salt-specific effects in protein solutions.

A seminal study in this direction was contributed by Taratuta
et al. \cite{Taratuta1990}. These authors investigated the
dependence of cloud-point temperature on the ionic strength of
the sodium phosphate buffer (figure~2 of that paper). In the
range of $I_\textrm{buffer}$ from $0.3$ to {$0.6$~mol\hspace{2
pt}dm$^{-3}$} and keeping {$\text{pH}=6.8$} constant, they found
no change in cloud-point temperature upon further increase of
the ionic strength of the buffer. It appears that due to strong
electrostatic screening, the Coulomb part of the
protein-protein interaction does not change upon further
increase of the buffer concentration. In continuation of this
interesting work, Taratuta and coworkers added
low-molecular-mass salts to the buffer, at the same time
decreasing the buffer content to keep the total ionic strength
equal to \mbox{$0.6$ mol\hspace{2pt}dm$^{-3}$}. In other words, they
varied the composition of the low-molecular-mass electrolyte,
keeping its total ionic strength constant. In the present
experimental study we follow this approach to examine the effects
of the added low-molecular-mass salts on the stability of protein
solutions.

Recently \cite{Kastelic2015} we proposed a new approach to
analyze the cloud-point temperature measurements. We modelled
protein molecules as hard spheres, with a number of
square-well attractive sites located on the surface. To obtain
measurable quantities we applied the thermodynamic perturbation
theory developed by
Wertheim \cite{Wertheim1986a,Wertheim1986b}. The approach was
used to analyze experimental data for $T_\textrm{cloud}$ in lysozyme
solutions. The calculations provided good fits to the
cloud-point curves of lysozyme in buffer-salt mixtures as a
function of the type and concentration of salt. In a spirit of
the chemical engineering theories, the approach was capable of
predicting full  coexistence curves, osmotic compressibilities and
second virial coefficients within the domain of concentrations
where $T_\textrm{cloud}$ were measured.

The work presented here is a continuation of our
previous study \cite{Kastelic2015} with one major difference
that we analyze our own $T_\textrm{cloud}$ measurements performed recently.
The review of literature revealed that experimental studies of
salt-specific effects are rarely systematic: sometimes salts
are added to protein solution in addition to buffer, sometimes
alone, forming the protein-salt mixture. Further, for the chosen
experimental method, the data collected in different
laboratories may scatter much more than it is suggested by
the precision of a single measurement. One reason for this lies
in the details of protein solution preparation, which appears to be
more important than it is actually recognized by most of the researchers. The
protein solutions are prone to ``age'' and one can obtain
different results with the freshly prepared or a few weeks old
samples. Such differences can be seen even within a single
paper \cite{Taratuta1990}.

Taking all these into account, and to avoid possible
experimental inconsistencies, we decided to perform our own
$T_\textrm{cloud}$ measurements on the well characterized solutions.
The data were taken on lysozyme samples purchased from a single
producer (Merck, Germany). We took all the necessary
precautions in preparing the solutions, for details see the
experimental part of the manuscript, to ensure a consistency of
the results and a fair comparison with theory. The measurements
were analyzed using the one-component model published
recently \cite{Kastelic2015}. Based on this analysis and
on our new $T_\textrm{cloud}$ measurements, we predicted other
thermodynamic quantities, including the full binodal curves and
osmotic second virial coefficients for lysozyme in phosphate
and acetate buffers in presence of low-molecular-mass salts.


\section{Experimental details}


\subsection{Materials and solution preparations}

Hen egg white lysozyme \mbox{($M_\textrm{2} = 14.388$~g mol$^{-1}$)} was
purchased from Merck Milipore, product number \mbox{105281}, lot
\mbox{K46535581 514}. The alkali metal salts ($>99$\%, KCl, NaCl,
KBr, NaBr, NaI, NaNO$_3$, NaH$_2$PO$_4\cdot$2H$_2$O, and
Na$_2$HPO$_4$) were obtained from Merck Milipore as well, while
CH$_3$CH$_2$COONa and NaSCN were obtained from Sigma Aldrich.
The first step was preparation of the lysozyme-buffer and
salt-buffer stock solutions. Dialyzing buffer was
\mbox{NaH$_2$PO$_4$/Na$_2$HPO$_4$} with ionic strength of \mbox{0.1
mol\hspace{2pt}dm$^{-3}$} and \mbox{$\text{pH}=6.8$}. Lysozyme was
dissolved in buffer and dialyzed against it for 24~h, using the
Spectra/Por Membrane$^{\circledR}$ dialysis membrane with
the $M_\textrm{w}$ cutoff of 3500~Da. During this time, the
buffer was changed three times. Concentrations of protein and
salts in stock solution were two times higher than in the
solution used in $T_\textrm{cloud}$ measurements. As often for mixed electrolytes,
the salt and buffer amounts are  given in ionic
strength, $I =  {\frac{1}{2}} \sum_i c_i z_i^2$, where sum goes
over all ionic species of salt, $i$ of concentration $c_i$
and electrovalence $z_i$. For $+1$:$-1$
salt $I$ is equal to its concentration, $c$ in
mol\hspace{2pt}dm$^{-3}$.

The low-molecular-mass salts were in presence of P$_2$O$_5$
dried for two hours at $T=130^\circ$C. Stock salt-buffer
solutions were prepared in such a way that solid components were weighted
and then filled
with distilled water in a filling flask up to the mark.
Mixtures of salts and protein were prepared
just before the measurements. The lysozyme concentration was
determined by measuring the absorbance at \mbox{$\lambda = 280$~nm}
and $25^\circ$C using a Cary 100 Bio (Varian)
spectrophotometer, which uses the Peltier block for temperature
regulation. The same instrument was used for the $T_\textrm{cloud}$
determination. The extinction coefficient of lysozyme was
\mbox{$2.635$ dm$^3$g$^{-1}$cm$^{-1}$} at $25^\circ$C. $\text{pH}$
was measured using the Iskra pH meter model MA5740 (Ljubljana,
Slovenia), using combined glass micro-electrode of type InLab
423 from Mettler Toledo (Schwerzenbach, Switzerland).
$\text{pH}$ of solutions were determined at the beginning and at the end of
the experiment. The deviations from the desired $\text{pH}$
values were always within $\pm0.1$.


\subsection{Cloud-point temperature measurements}

$T_\textrm{cloud}$ is defined as the temperature where upon cooling
the first opacification is noticed in solution under study. The
cloudiness was in our case detected by an increase in the solution
absorbance at wavelength \mbox{$\lambda = 340$~nm}.
As noticed before, the measured cloud-point temperatures may
depend on the cooling rate \cite{Grigsby2001}. In an initial
investigation of the system, we measured $T_\textrm{cloud}$ at three
different cooling rates: 0.1, 0.5, and \mbox{$1.0^\circ$C\;min$^{-1}$}
and extrapolated these results to cooling rate equal to zero. In
the $T_\textrm{cloud}$ measurements reported here, we used the cooling
rate equal to \mbox{$0.1^\circ$C\;min$^{-1}$}, which yields the results
very close (within $\pm 0.1^\circ$C) to the extrapolated value.
Reversibility of the process was verified by warming up the
sample above the $T_\textrm{cloud}$ and by cooling it again to repeat
the measurement. Like some other authors before us, we measured
both the $T_\textrm{cloud}$ and $T_\textrm{clear}$; the latter is the
temperature where the solution becomes clear again. The differences
between these two temperatures were in the range from 1.0 to
\mbox{$4.0^\circ$C}, depending on the salt type and concentration.
While other researchers \cite{Grigsby2001} take
the average of $T_\textrm{cloud}$ and $T_\textrm{clear}$ as a final value, we
report the actual $T_\textrm{cloud}$ values in our results. In view of
the observed differences between $T_\textrm{cloud}$ and $T_\textrm{clear}$
values, the absolute error in temperature of cloud-point
determination is estimated to be between {$\pm1.0$} and {$\pm 2.0^\circ$C}.


\section{Theoretical part}

The theoretical model used in this study \cite{Kastelic2015} is
based on the observation that the range and directionality of the
attractive interactions between protein molecules determine
their phase
behaviour \cite{Frenkel1997,Lomakin1996,Lomakin1999,Sear1999,Rosch2007}.
Previous studies suggested the  appearance of a liquid-liquid
coexistence region, which turns out to be meta-stable with
respect to the
solidification \cite{Sear1999,Broide1991,Rosenbaum1996,Muschol1997}.
This is in contrast with the behavior of the systems composed
of van der Waals type of particles, where the range of
interaction between molecules is comparable with their size.
Theoretical methods suitable to the study of systems of molecules
interacting with strong directional forces have been proposed
by Wertheim \cite{Wertheim1986a,Wertheim1986b} and further
developed by many other authors \cite{Kalyuzhnyi2007,Chapman1988}. The
one-component model of
protein solution, which in some aspects resembles simple water
models \cite{Kolafa1987,Bizjak2009}, has recently been used
to analyze the experimental data for phase diagrams of lysozyme and
$\gamma$-crystallin solutions \cite{Kastelic2015}. For
convenience of a reader, the descriptions of the model and
theory are briefly repeated below.

\begin{figure}[!t]
\centering
\includegraphics[keepaspectratio=true,width=9cm]{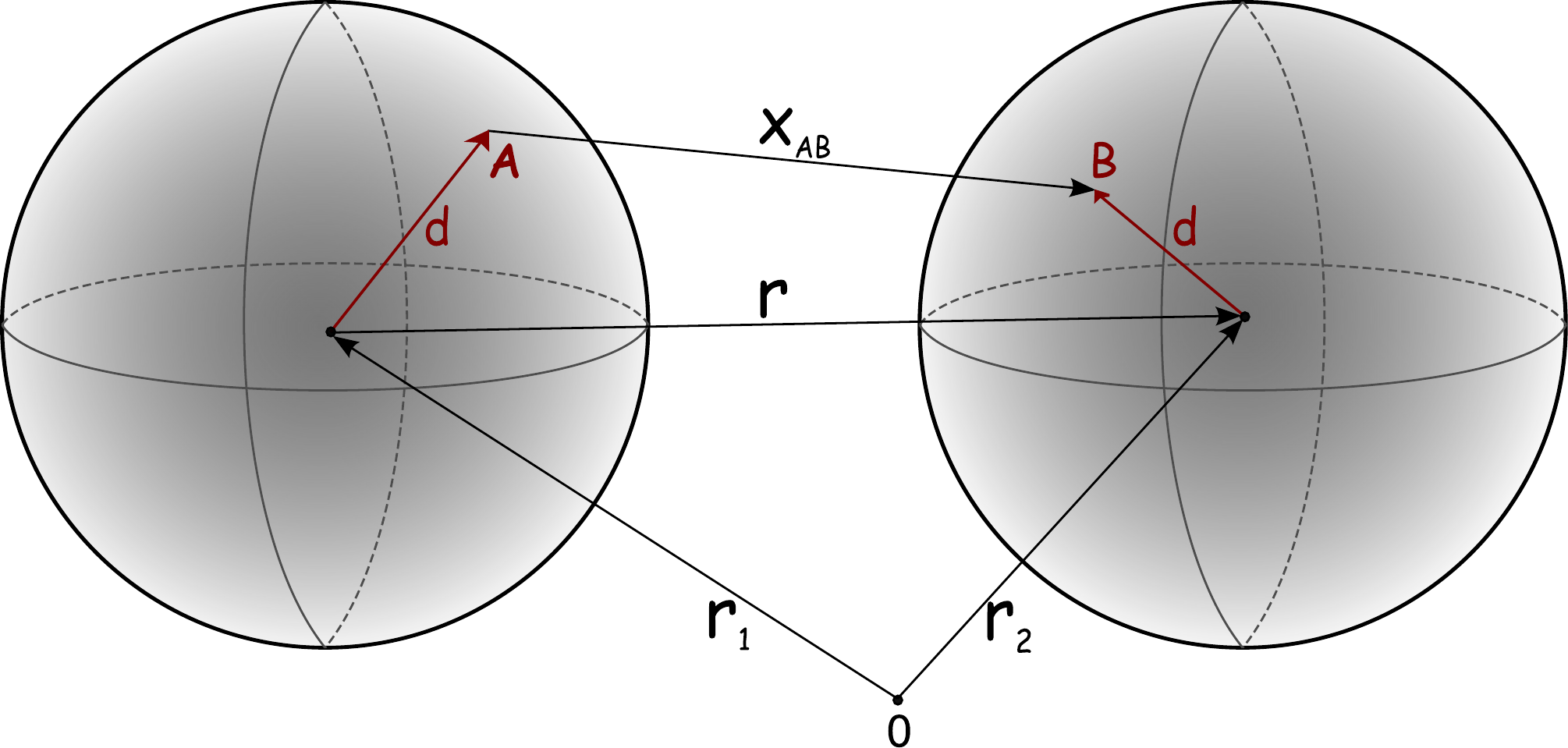}
\caption{(Color online) The pair potential consists of the hard-sphere and site-site
contributions. Only one site~--- of the set of $M$ sites~--- is shown on each
sphere. \label{potencial}}
\end{figure}

We model the solution as a system of $N$ protein molecules with
number density $\rho=N/V$ at temperature $T$ and volume $V$.
Protein molecule is pictured as a sphere of diameter $\sigma$
with the attractive square-well sites through which it
interacts with other protein molecules. The solution is treated
as a quasi one-component system, where the solvent (water,
buffer, and low-molecular-mass salt) merely modifies the
interaction between solutes. We assume the protein-protein
pair potential to be composed of: (i) the hard-sphere part
$u_\textrm R(r)$ and (ii) attractive contributions, $u_\textrm{AB}$, caused by
the (short-range) square-well sites localized on the surface
of the protein \cite{Wertheim1986a}
\begin{eqnarray}
u(\mathbf{{r}})=u_\textrm R(r)+\sum_{\textrm A\in\Gamma}
\sum_{\textrm B\in \Gamma}u_\textrm {AB}(\mathbf{{x}}_\textrm {AB}).
\end{eqnarray}
In this expression, $\mathbf{{r}}$
($r=|\mathbf{{r}}|$) is the vector between the centers
of molecules, $\mathbf{{x}}_\textrm {AB}$ is the vector
connecting sites A and B on two different protein molecules and
$\Gamma$ denotes the set of sites, see figure~\ref{potencial}.
We examine the case where $M$ equal sites are distributed over
the surface of the spherical protein; in other words, the
displacement length  $d$ is $0.5\sigma$. The pairwise additive
potential is then written as follows:
\begin{eqnarray}
u_\textrm R(r) & = & \left\{\begin{array}{ll}
\infty, & \qquad \text{for} \quad r < \sigma,\\
0, & \qquad \text{for} \quad r \geqslant \sigma,
\end{array} \right. \\
u_\textrm {AB}(\mathbf{{x}}_\textrm {AB}) & = &
\left\{\begin{array}{ll}
-\varepsilon_\textrm {W}, & \qquad \text{for} \quad |\mathbf{{x}}_\textrm {AB}| < a_\textrm {W}, \\
0, & \qquad \text{for} \quad |\mathbf{{x}}_\textrm {AB}| \geqslant a_\textrm {W}.
\end{array} \right.
\end{eqnarray}
Here, $\varepsilon_\textrm {W}$ ($>0$) is the square-well potential
depth and $a_\textrm {W}$ is its range. The interaction between the sites
is only effective for the site-site distance
$|\mathbf{{x}}_\textrm {AB}|$ being smaller than $a_\textrm {W}$. The
multiple site bonding is prevented by applying the
condition \cite{Wertheim1986a,Wertheim1986}
\begin{eqnarray}
0 <a_\textrm {W} < \sigma-\sqrt{3}d.
\end{eqnarray}
As usually in such studies, the additivity of the free energy
terms is assumed
\begin{eqnarray}
A = A^\textrm {id} + A^\textrm {hs} + A^\textrm {ass}, \label{free}
\end{eqnarray}
where $A^\textrm {id}$ is the ideal part \cite{Hansen_McDonnald2006},
$A^\textrm {hs}$ is the hard-sphere part \cite{Mansori1971}, while
$A^\textrm {ass}$ stands for the site-site association
contribution \cite{Wertheim1986a,Wertheim1986b,Chapman1988}
\begin{eqnarray}
\frac{\beta A^\textrm {ass}}{N} = M\bigg(\ln{X}-\frac{X}{2}+\frac{1}{2}\bigg),
\label{ass}
\end{eqnarray}
where $\beta=(k_\textrm BT)^{-1}$ and $k_\textrm B$ is Boltzmann's constant.
The association parameter $X$ defines the average fraction of
the molecules not bonded to any site \cite{Chapman1988}
\begin{eqnarray}
X=\frac{1}{1 + MX\rho\Delta_\textrm {AB}}. \label{MAL}
\end{eqnarray}
Further, the $\Delta_\textrm {AB}$ term is related to the hard-sphere
fluid through the radial distribution function $g^\textrm {hs}(r)$ via
the expression \cite{Wertheim1986}
\begin{eqnarray}
\Delta_\textrm {AB} = 4\pi g^\textrm {hs}(\sigma)\int_{\sigma}^{2d+a_\textrm {W}}
\bar{f}_\textrm {ass}(r)r^2\textrm{d}r. \label{Delta}
\end{eqnarray}
The radial distribution function $g^\textrm {hs}(r)$ is calculated by
the Ornstein-Zernike integral equation theory using the
Percus-Yevick (PY) closure \cite{Hansen_McDonnald2006}, yielding
\begin{eqnarray}
g^\textrm {hs}(\sigma) & = & \frac{2+\eta}{2(1-\eta)^2}, \label{g_contact}
\end{eqnarray}
where $\eta=\pi\rho\sigma^3/6$ is the packing fraction of hard
spheres. Further, $\bar{f}_\textrm {ass}(r)$, is the angular average of the
Mayer function, obtained analytically \cite{Wertheim1986}
\begin{eqnarray}
\bar{f}_\textrm {ass}(r) & = & \frac{\exp{(\beta
\varepsilon_\textrm {W})}-1}{24d^2r}(a_\textrm W+2d-r)^2(2a_\textrm W-2d+r).
\end{eqnarray}
Once the Helmholtz free energy, equation (\ref{free}), is known,
other thermodynamic quantities, among them the osmotic pressure
${\Pi}$ and chemical potential of the protein species, ${\mu}$,
can be calculated. By using equations (\ref{ass}), (\ref{MAL}) and
(\ref{g_contact}), we get expressions for osmotic pressure and then for the
chemical potential
\begin{eqnarray}
\beta \Pi^\textrm {ass} & = &
-\frac{(M\rho)^2\Delta_\textrm {AB}X(2-X)(1+2\eta)}{(1+2M\rho\Delta_\textrm {AB}X)(1-\eta)(
2+\eta)}, \label{Maxwell_P} \\
\mu & = & \frac{A}{N}+\frac{\Pi}{\rho}. \label{Maxwell_mu}
\end{eqnarray}
Ideal and hard sphere contributions to the free energy and pressure can be
found elsewhere \cite{Hansen_McDonnald2006}. At this step we can calculate
the cloud point temperature, as well as the whole liquid-liquid
coexistence curve, by applying the Maxwell construction. Another important
theoretical and experimental quantity is the
second virial coefficient, $B_\textrm 2$, defined as:
\begin{equation}
{\frac{\Pi}{ k_\textrm B T}} = \rho  + B_\textrm 2 \rho^2  + \ldots \; .
\label{pi}
\end{equation}
We calculated this quantity as suggested by Bianchi et
al. \cite{Bianchi2006}
\begin{eqnarray}
B_2 = B_2^\textrm{(hs)}-2\pi
M^2\int_{\sigma}^{2d+a_\textrm {W}}\hspace{-0.25cm}\bar{f}_\textrm {ass}(r)r^2\textrm{d}r.
\label{B2_wert}
\end{eqnarray}
Here, $B_\textrm 2^\textrm{(hs)}=2 \pi \sigma^3/3$ is the second virial
coefficient of hard spheres. Note that the integral in equations (\ref{Delta})
and (\ref{B2_wert}) can be calculated analytically.
Model parameters used in calculations are given in table~\ref{Table1}.
\begin{table}[!h]
\caption{Model parameters used in the calculations. $M_{2}$ is the
molar mass of the lysozyme.\label{Table1}}
\vspace{-2ex}
\begin{center}
\begin{tabular}{ll}
\multicolumn{2}{c}{} \\
\hline
\hline
$\sigma$\;[nm] & 3.430 \\
$d$\;[nm] & 1.715 \\
$a_\textrm W$\;[nm] & 0.180 \\
$M_\textrm{2}$\;[g\hspace{2 pt}mol$^{-1}$] & 14388 \\
$M$ & 10 \\
\hline
\hline
\end{tabular}
\end{center}
\end{table}

\vspace{-3mm}

\section{Results and discussion}


\subsection{Cloud-point temperatures for lysozyme-buffer-salt mixtures}

\begin{figure}[!b]
\centering
\includegraphics[keepaspectratio=true,scale=0.35]{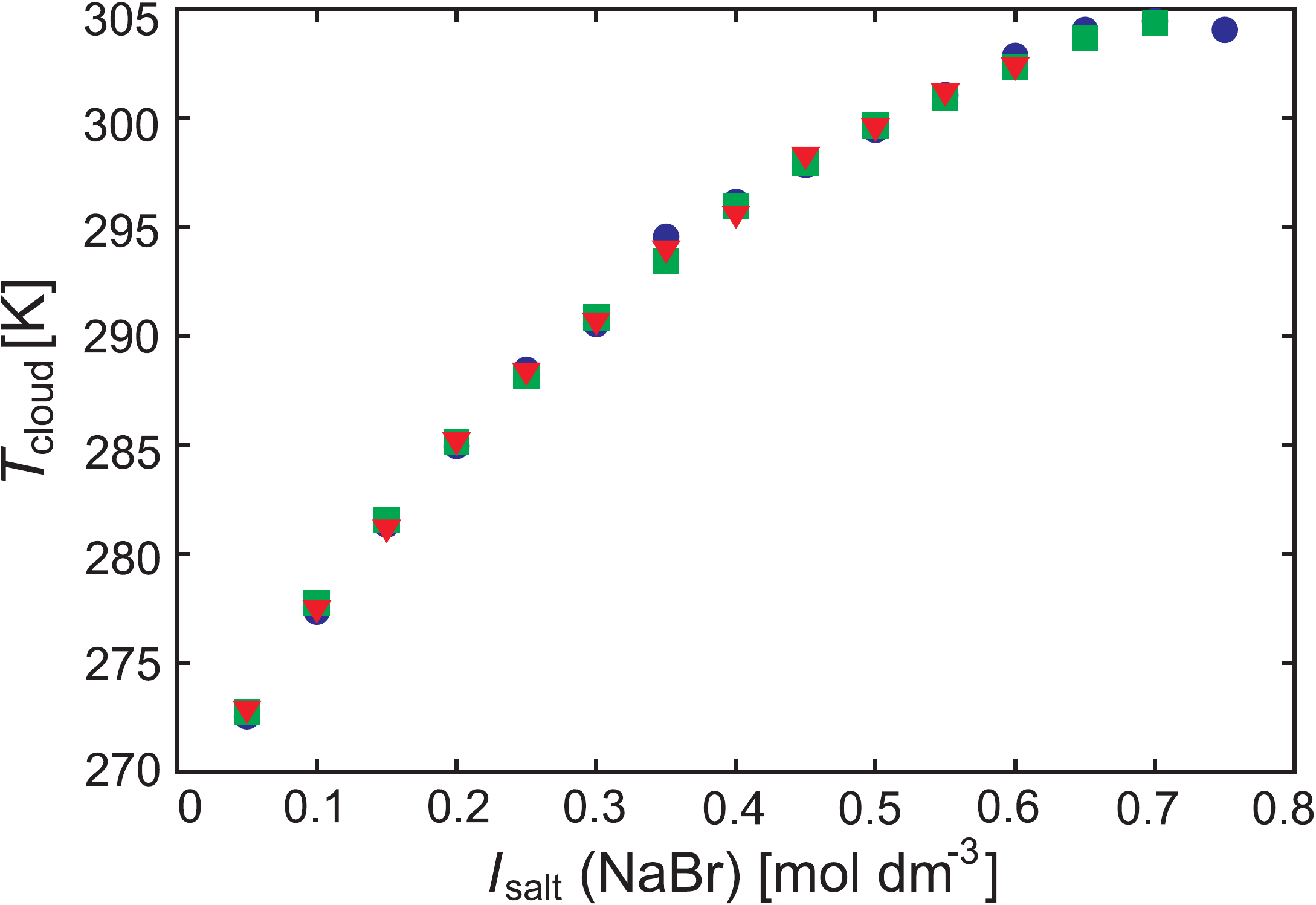}
\caption{(Color online) Experimental results for the cloud-point temperatures,
$T_\textrm {cloud}$, in the protein-phosphate buffer mixtures as a function of the
concentration of the added NaBr (${I_\textrm {salt}}$). Experimental
conditions: total ionic strength is \mbox{$I_\textrm {total} =
0.7$} (\textcolor{red}{$\blacktriangledown$}), 0.8
(\textcolor{green}{\scriptsize$\blacksquare$}), and 0.9
(\textcolor{blue}{\Large$\bullet$}) \mbox{mol\hspace{2
pt}dm$^{-3}$}, $\text{pH}=6.8$ and
protein concentration $\gamma=90$ \mbox{g\hspace{2 pt}dm$^{-3}$}.}
\label{buffer}
\end{figure}

Taratuta et al. \cite{Taratuta1990} noticed that after a
sufficient amount of buffer is added, $T_\textrm {cloud}$
becomes insensitive to a further increase of ionic strength.
This observation suggests that at certain ionic strength, the
Coulomb interaction between proteins
becomes sufficiently screened. We confirmed this finding for
lysozyme in mixture with phosphate buffer
\mbox{($\text{pH}=6.8$)} and NaBr (see figure \ref{buffer}).
In this graph we present $T_\textrm {cloud}$ taken as a
function of the ionic strength of the added sodium bromide
($I_\textrm {salt}\equiv c_\textrm {salt}$) at a constant total
ionic strength \mbox{$I_\textrm {total} = I_\textrm {salt} +
I_\textrm {buffer}$}, with $I_\textrm {total}$ equal to 0.7,
0.8, and \mbox{0.9 mol\hspace{2pt}dm$^{-3}$}, respectively.
Symbols representing the measurements at different $I_\textrm
{total}$ fall --- within the experimental error --- on the same
curve. These results indicate that for a given value of
$I_\textrm {salt} \neq 0$, $T_\textrm {cloud}$ is insensitive
to the $I_\textrm {total}$ variations above 0.6
mol\hspace{2pt}dm$^{-3}$. Due to the experimental limitations
of our apparatus, no $T_\textrm{cloud}$ values could be
determined below $-6^\circ$C.

\begin{figure}[!t]
\centering
\includegraphics[keepaspectratio=true,scale=0.35]{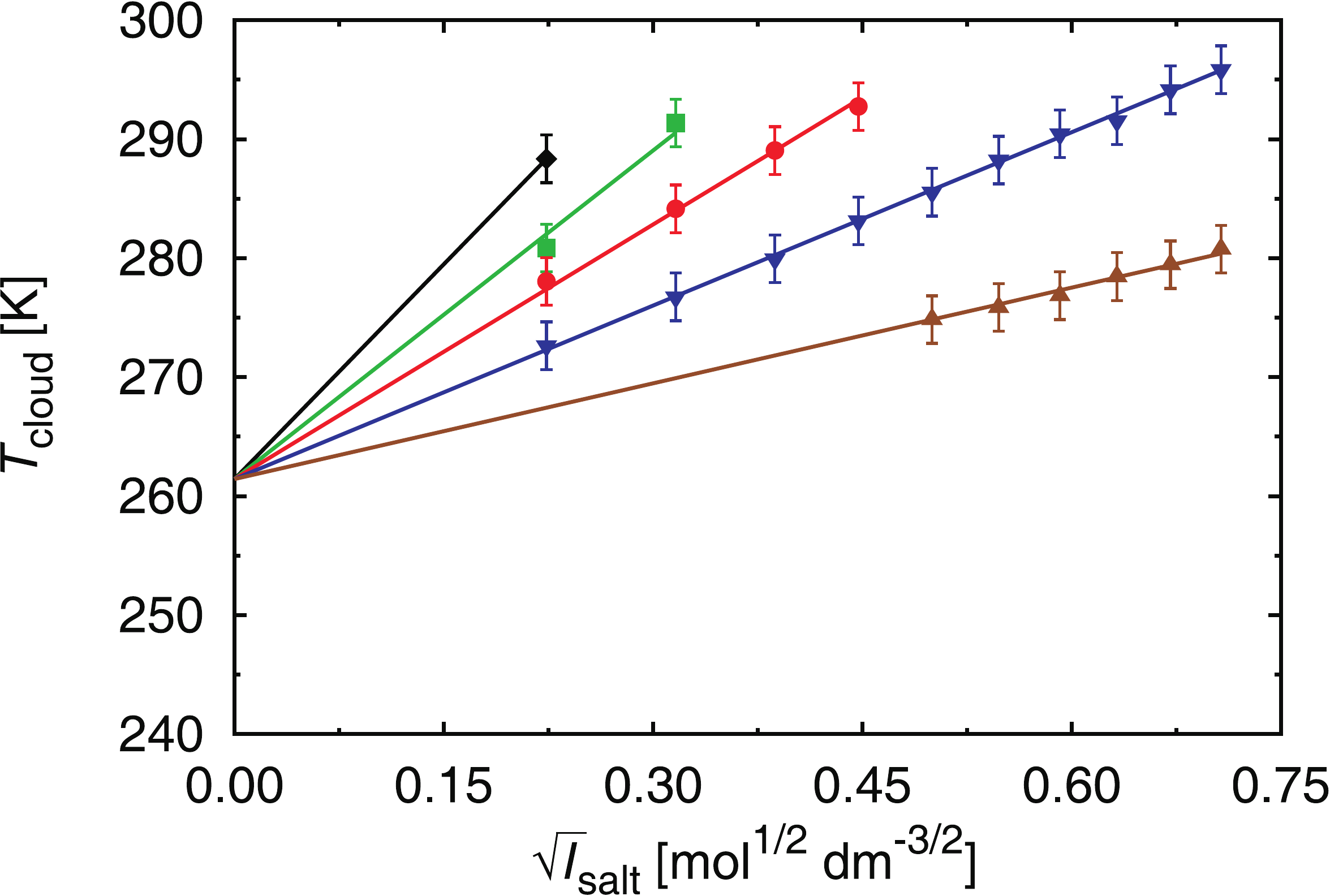}
\caption{(Color online) Experimental results (symbols) for the cloud-point temperatures,
$T_\textrm{cloud}$, in the protein-phosphate buffer mixtures as a function of the
square root of simple salt content, $\sqrt{I_\textrm{salt}}$. Experimental
conditions: \mbox{$I=0.6$ mol\hspace{2 pt}dm$^{-3}$}, \mbox{$\text{pH}=6.8$} and
protein concentration
\mbox{$\gamma=90$ g\hspace{2 pt}dm$^{-3}$}.
Legend: NaSCN (\textcolor{black}{$\blacklozenge$}), NaI (\textcolor{green}{\scriptsize$\blacksquare$}),
NaNO$_3$ (\textcolor{red}{\Large$\bullet$}), NaBr
(\textcolor{blue}{$\blacktriangledown$}), and NaCl (\textcolor{brown}{$\blacktriangle$}).
Theoretical predictions (lines) according to equation (\ref{Correlation-eq}) and
table~\ref{Correlation}.}
\label{Tc-phosphate}
\end{figure}

\begin{figure}[!b]
\centering
\includegraphics[keepaspectratio=true,scale=0.35]{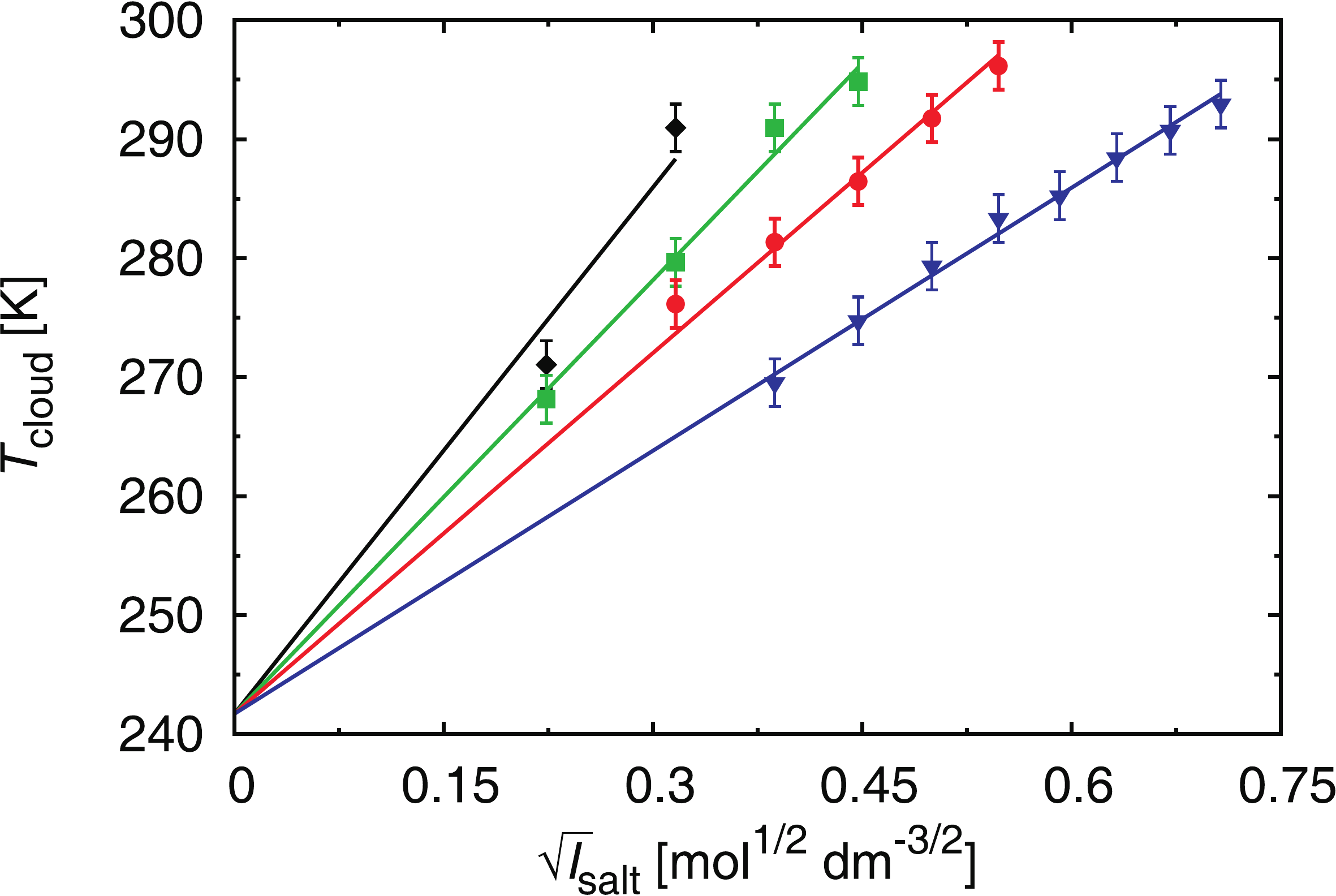}
\caption{(Color online) Experimental results (symbols) for the cloud-point temperatures, $T_\textrm{cloud}$, for various salts added to the protein-acetate buffer
mixture as a function of the square root of simple salt content, $\sqrt{I_\textrm{salt}}$.
Legend as for figure~\ref{Tc-phosphate}, but instead of NaBr there is KBr in this figure. Theoretical predictions according to equation (\ref{Correlation-eq}) and table~\ref{Correlation-ac} with the $M=10$ (lines). Experimental conditions as for the previous figure except for $\text{pH}$, which
is equal to 4.6.} \label{Tc-acetate}
\end{figure}

In figure \ref{Tc-phosphate}, the experimental results for
$T_\textrm{cloud}$ at \mbox{$\text{pH}=6.8$}, lysozyme concentration
\mbox{$\gamma=90$~g\hspace{2 pt}dm$^{-3}$}, and total ionic strength
\mbox{($I_\textrm{buffer}$ + $I_\textrm{salt}$)} equal to \mbox{0.6~mol\hspace{2
pt}dm$^{-3}$} are shown. Analogous results for acetate
buffer are shown in figure~\ref{Tc-acetate}. The experiments suggest a
square root functional dependence between the $T_\textrm{cloud}$ and ionic
strength of the added electrolyte. Notice that the $T_\textrm{cloud}$ at
{$I_\textrm{salt} = 0$} should be the same for all the salts. The extrapolated
value of $T_\textrm{cloud}$ to {$I_\textrm{salt}= 0$}, as we have already mentioned, this
point is experimentally not accessible, is {$-12 \pm 2^\circ$C}.
The salt-specific effects in $T_\textrm{cloud }$ measurements have
been observed in several previous experimental
papers \cite{Taratuta1990,Zhang2009}. An increase of the
cloud-point temperature can be interpreted as a decrease of
stability of the system. Considering that the total ionic
strength is constant for all the samples studied in figure~\ref{Tc-phosphate}, this instability has been ascribed to the
salt adsorption occurring at the protein
surface \cite{Taratuta1990}. Lysozyme solutions at
\mbox{$\text{pH}=6.8$} assume a positive net charge. That is why
the effects of anions are strong.

In our model, the strength of the protein-protein attraction can
only be regulated by the depth of the attractive square-well
potential, $\epsilon_\textrm{W}$. In view of the new experimental
results, shown in figure~\ref{Tc-phosphate}, we suggest this
quantity to be correlated with the ionic strength of the added
low-molecular-mass salt ($I_\textrm{salt}$) as follows:
\begin{eqnarray} \epsilon_\textrm{W}(I_\textrm{salt})/k_\textrm B &=&
a \, \sqrt I_\textrm {salt} + b. \label{Correlation-eq}
\end{eqnarray}
The parameters of this equation, leading to a good agreement with
experimental data for solutions in phosphate buffer, are given
in table~\ref{Correlation}. As we see, the slope of equation
(\ref{Correlation-eq}) (parameter $a$) varies from salt to salt.

\begin{table}[!h]
\caption{Parameters $a$ and $b$ defining equation
(\ref{Correlation-eq})~--- phosphate buffer \mbox{($\text{pH}=6.8$)},
protein concentration \mbox{$\gamma = 90$ g\hspace{2
pt}dm$^{-3}$}.\label{Correlation}}
\vspace{2ex}
\begin{center}
\begin{tabular}{l|c|c}
\hline\hline
 & $a$ [K\hspace{2 pt}dm$^{3/2}$\hspace{2 pt}mol$^{-1/2}$] & $b$ [K] \\
\hline\hline
NaSCN & 1055 & \multirow{5}{*}{2293} \\
\cline{1-2}
NaI & 807 &  \\
\cline{1-2}
NaNO$_3$ & 625 & \\
\cline{1-2}
NaBr & 426 & \\
\cline{1-2}
NaCl & 235 & \\
\hline\hline
\end{tabular}
\end{center}
\end{table}

\begin{figure}[!t] \centering
\includegraphics[keepaspectratio=true,scale=0.35]{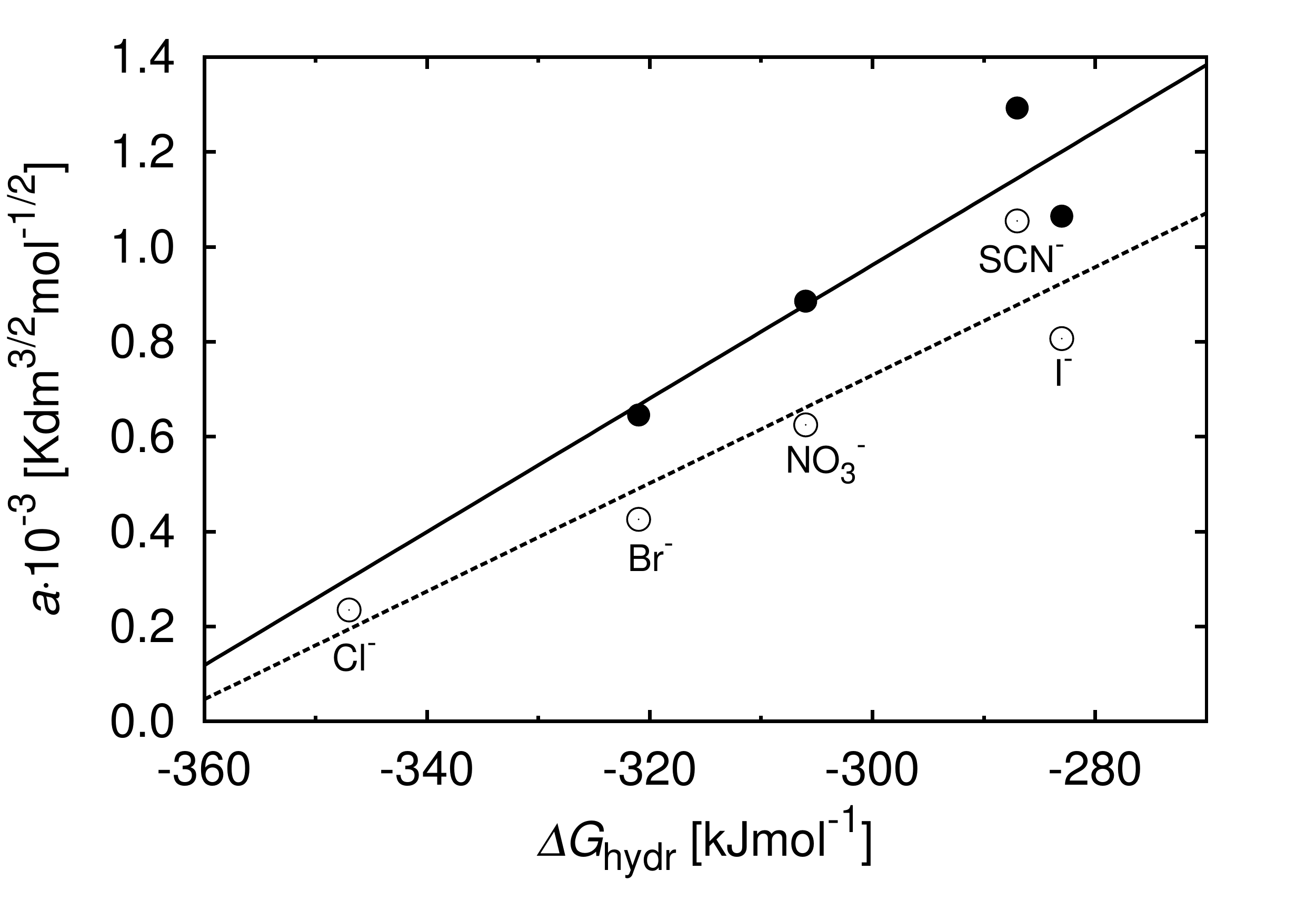}
\caption{Correlation of the slope $a$ of equation (\ref{Correlation-eq}) with the
hydration Gibbs free energies $\Delta G_\textrm{hydr}$ \cite{Marcus1997}
of the corresponding anions. The lines are the best least-square fit through the data. The
upper curve belongs to $\text{pH}= 4.6$ and
the lower one belongs to $\text{pH}= 6.8$.}
\label{Correlation-G}
\end{figure}

The parameters leading to a good agreement with experimental data
for acetate buffer, are given in table~\ref{Correlation-ac}.
The extrapolated value of $T_\textrm{cloud}$ at $I_\textrm{salt}=0$ is in
this case equal to $-31 \pm 2^\circ$C. It is of interest to correlate the slope in equation~(\ref{Correlation-eq}), which depends solely on the potential
well-depth, with the hydration free energy of the salt anion.
These results are shown in figure~\ref{Correlation-G} for two
different buffers~--- phosphate (lower curve) and acetate (upper curve).
As we see, the two lines show the same trend. The ion dependent
shift between the two lines seem to reflect the effects of the
buffers present in systems. We see that, as found before for
polyelectrolyte \cite{Serucnik2012} and lysozyme
solutions \cite{Kastelic2015,Boncina2010}, the strength of
protein-protein interaction is roughly correlated with the free
energy of counterion solvation $\Delta G_\textrm{hydr}$.

\begin{table}[!h]
\caption{Parameters $a$ and $b$ defining equation (\ref{Correlation-eq})~--- acetate buffer ($\text{pH}=4.6$), protein
concentration \mbox{$\gamma=90$ g\hspace{2
pt}dm$^{-3}$}.\label{Correlation-ac}}
\vspace{2ex}
\begin{center}
\begin{tabular}{l|c|c}
\hline
\hline
 & $a$ [K\hspace{2 pt}dm$^{3/2}$\hspace{2 pt}mol$^{-1/2}$] & $b$ [K] \\
\hline
\hline
NaSCN & 1293 & \multirow{4}{*}{2120} \\
\cline{1-2}
NaI & 1065 &  \\
\cline{1-2}
NaNO$_3$ & 886 & \\
\cline{1-2}
KBr & 646 & \\
\hline
\hline
\end{tabular}
\end{center}
\end{table}


\subsection{From $T_\text{cloud}$ to $B_{2}$ and liquid-liquid phase-diagram}

We can use experimental information collected in tables
(\ref{Correlation}) and (\ref{Correlation-ac}), to calculate
$\epsilon_\textrm{W}(I_\textrm{salt})$ from equation (\ref{Correlation-eq})
and then to predict full binodal curves under conditions
($I_\textrm{salt}$, $I_\textrm{total}$), where no such experiments have been
done yet.
These results are for NaBr and NaI in aqueous mixture of lysozyme shown
in our next figure~\ref{PD_evolution} in phosphate buffer,
$\text{pH}=6.8$ and in acetate buffer, $\text{pH}=4.6$ for KBr and NaI.
Notice that total ionic strength
($I_\textrm{buffer} + I_\textrm{salt}$) is constant and equal to \mbox{0.6
mol\hspace{2pt}dm$^{-3}$}, while $I_\textrm{salt} \equiv c_\textrm{salt}$
varies.

\begin{figure}[!b]
\begin{center}
\includegraphics[keepaspectratio=true,width=0.45\textwidth]{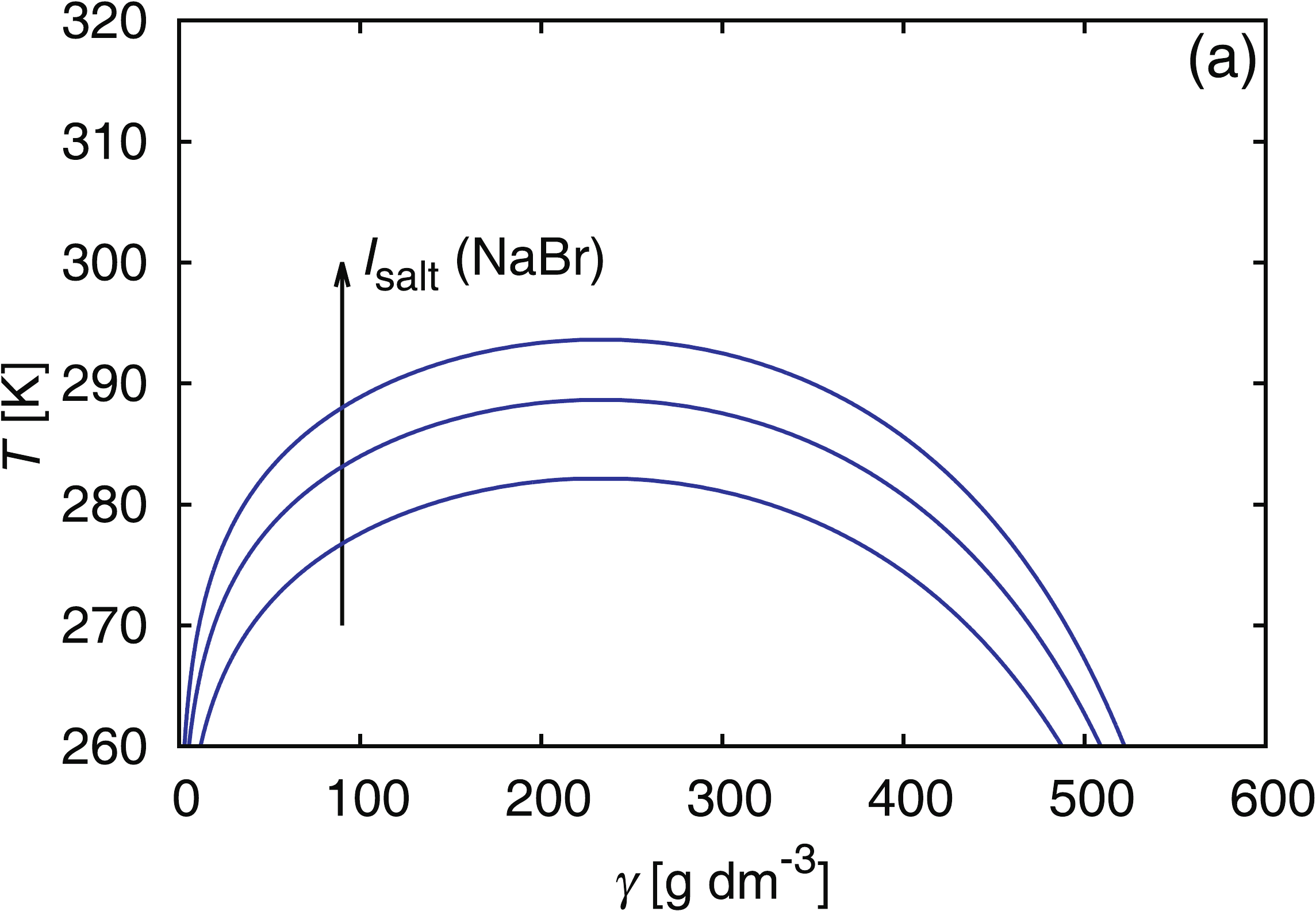}
\qquad
\includegraphics[keepaspectratio=true,width=0.45\textwidth]{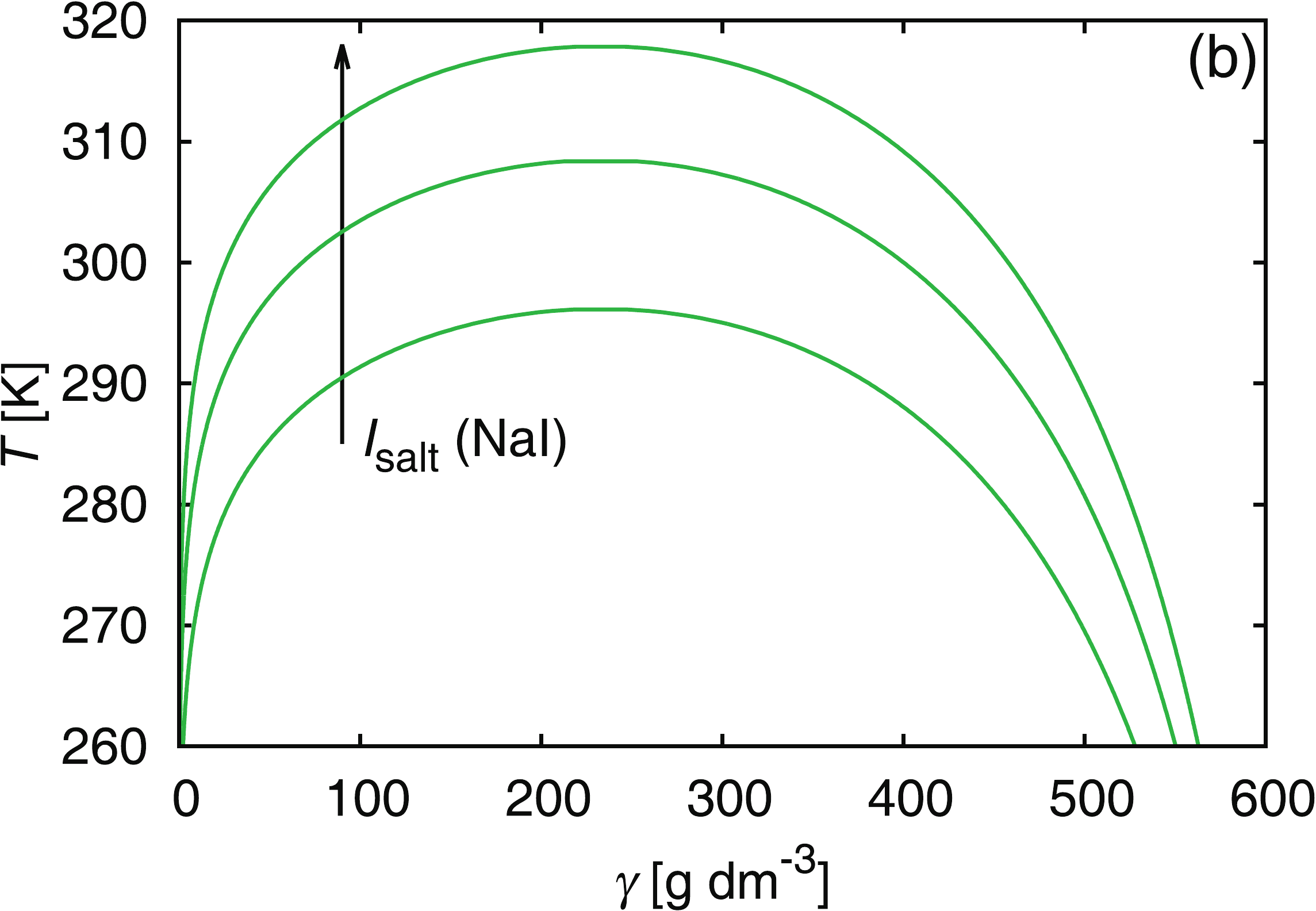} \\
\includegraphics[keepaspectratio=true,width=0.45\textwidth]{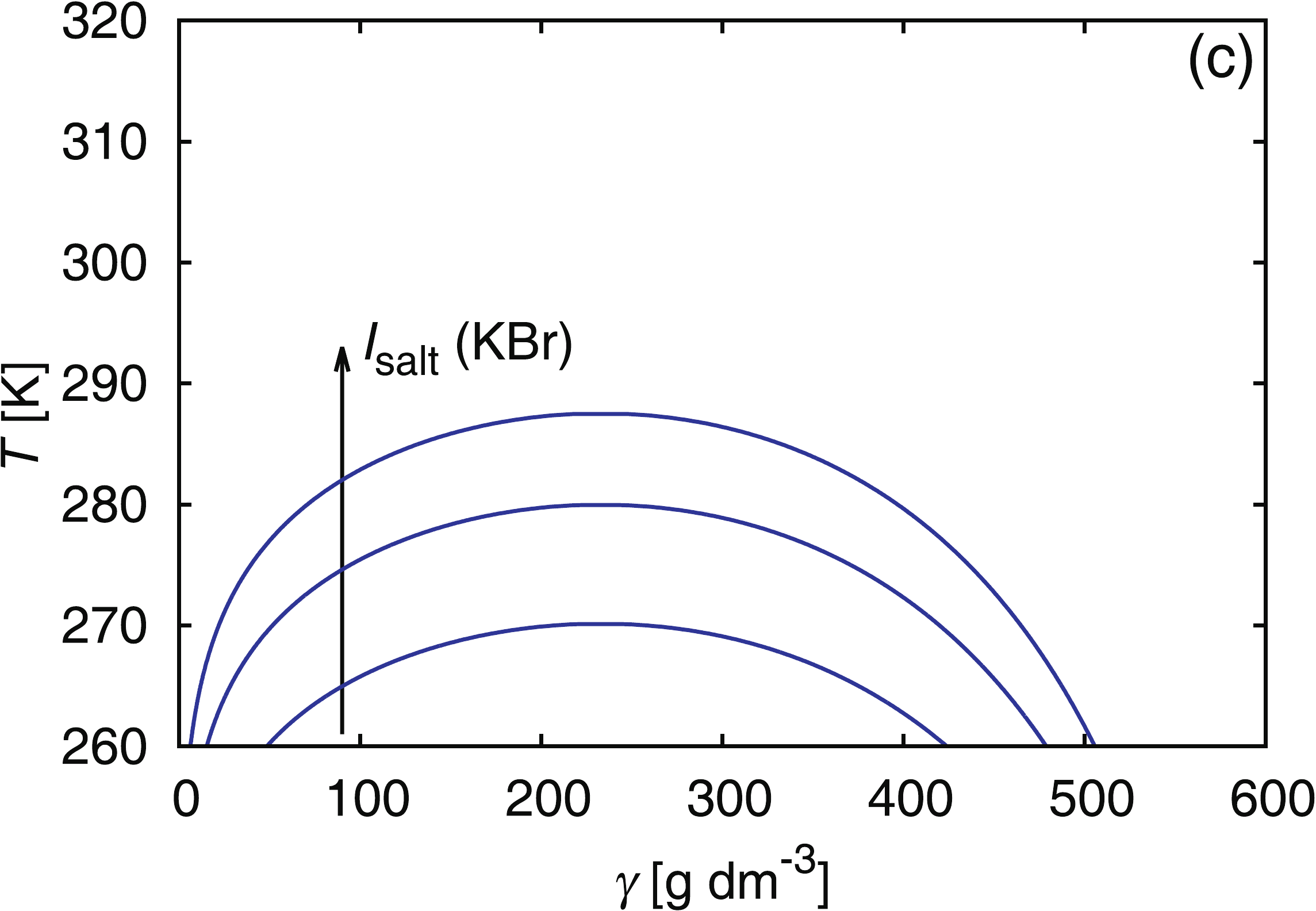}
\qquad
\includegraphics[keepaspectratio=true,width=0.45\textwidth]{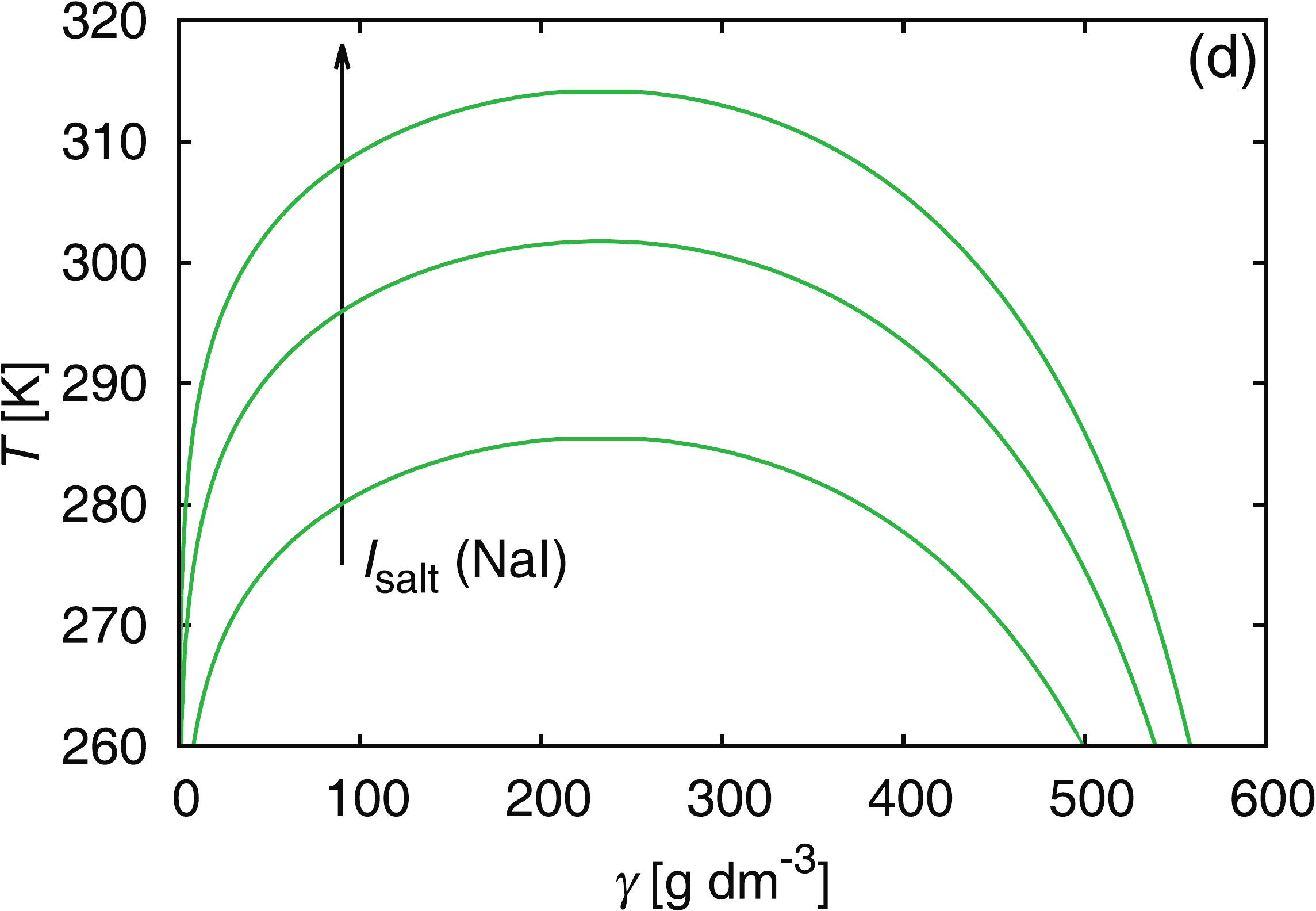}
\end{center}
\caption{(Color online)  Panels (a) and (b) phosphate buffer, panels (c) and (d) acetate
buffer. Coexistence curve shifts toward higher temperatures, when
$I_\textrm{salt}$  increases by the order 0.1, 0.2 and 0.3~mol\hspace{2
pt}dm$^{-3}$, as indicated by an arrow. Phase diagram evolution was
calculated through the Maxwell construction, using equations
(\ref{Maxwell_P}), (\ref{Maxwell_mu}) in conjunction with equation
(\ref{Correlation-eq}) and associated tables~\ref{Correlation} and
\ref{Correlation-ac}. Mass concentration of protein, $\gamma$, is related
to $\rho$ as $\gamma=\rho M_{2}/N_\mathrm{A}$, where $N_\mathrm{A}$ is
Avogadro's number.}
\label{PD_evolution}
\end{figure}
It is important to stress that for proteins, the liquid-liquid
boundary is located below the solid-liquid boundary,
indicating the meta-stability of such
systems \cite{Sear1999,Tavares2004}. When the saturated protein
solution is cooled, it may undergo liquid-liquid phase
transition before it actually crystalizes. This discriminates
proteins from most of low-molecular weight mixtures, where such
meta-stabilities were not observed. The liquid-liquid
coexistence curve can be determined
experimentally \cite{Broide1996,Zhang2009}.

\begin{figure}[!h]
\begin{center}
\includegraphics[width=0.48\textwidth]{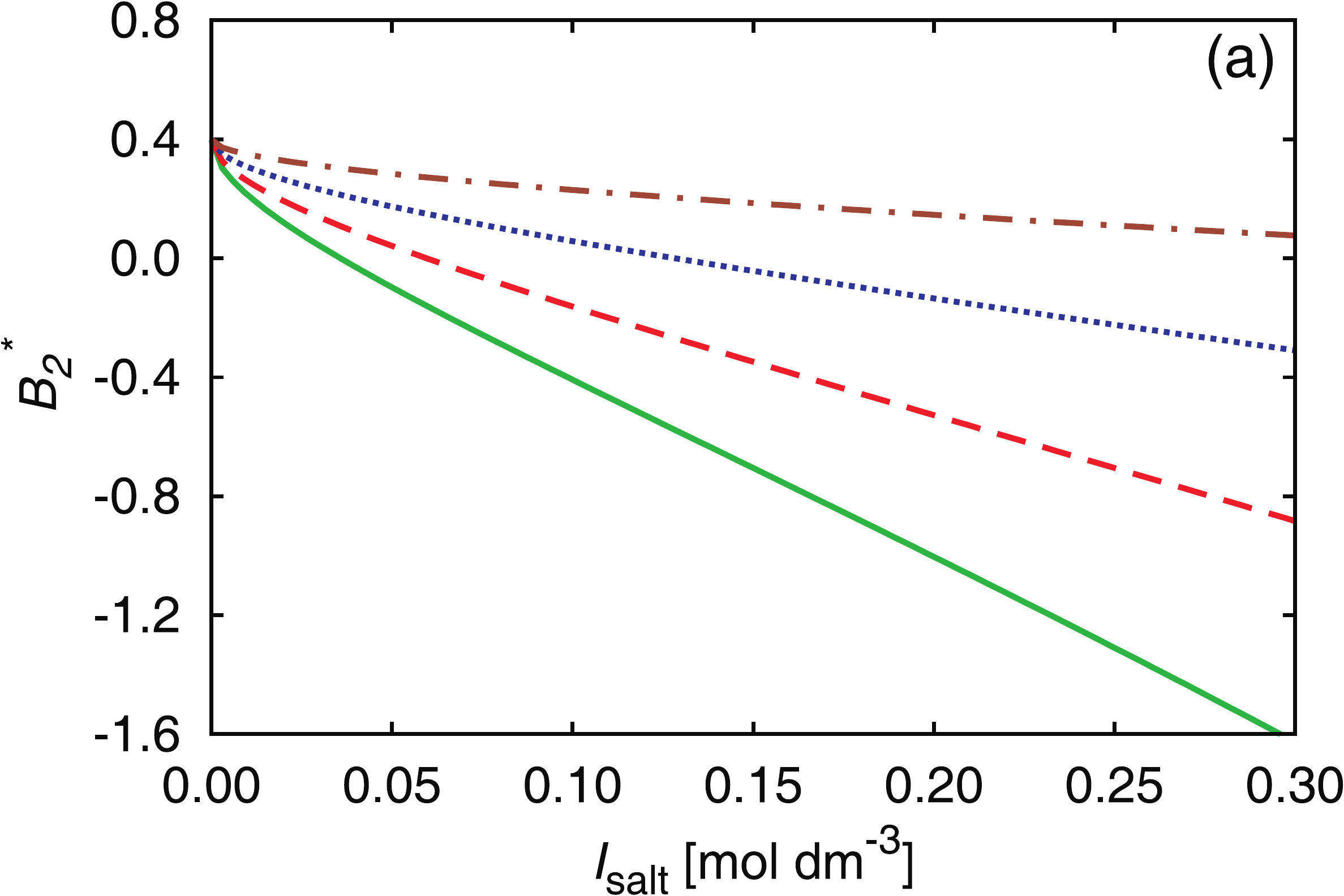} \quad
\includegraphics[width=0.48\textwidth]{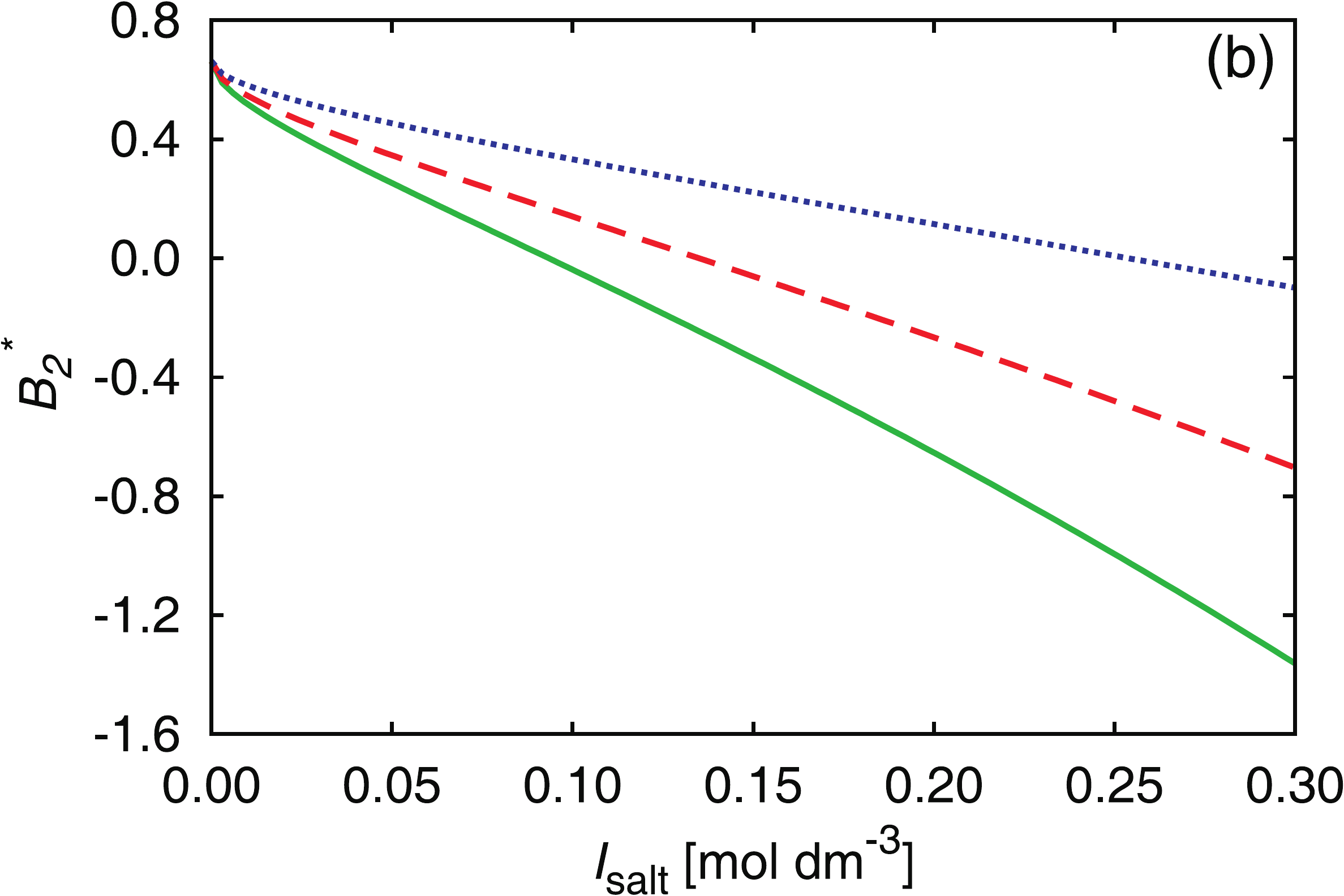}
\end{center}
\caption{(Color online)  The reduced second virial coefficient $B_\textrm 2^*=B_\textrm 2/B_\textrm 2^\textrm {(hs)}$ as calculated from equation (\ref{B2_wert}) in conjunction with equation (\ref{Correlation-eq}) at $T=300$~{K}. Parameters $a$ and $b$ are listed in tables~\ref{Correlation} and \ref{Correlation-ac} for respective buffers.
Left-hand panel (a): phosphate buffer, $\text{pH}=6.8$ (from the bottom to the top: NaI, NaNO$_3$, NaBr, NaCl).  Right-hand panel (b): acetate buffer, $\text{pH}=4.6$ (from the bottom to the top: NaI, NaNO$_3$, KBr).}
\label{B2-figs}
\end{figure}

Another quantity of interest is the second virial coefficient
$B_2$, a critical parameter in controlling the protein
aggregation. The latter process is of practical interest for
pharmaceutical industry \cite{Frokjaer2005,Valente2005}. Wilson
and co-workers \cite{George1994,George1997} discovered that in
order to grow well-defined crystals, the second virial
coefficient should be slightly negative.
%
%
The salt-specific effects can be observed also in figure~\ref{B2-figs}. For each of the two buffers (they determine the
$\text{pH}$ of solutions), the reduced second virial
coefficient decreases with increasing $I_\textrm{salt}$. The decrease
is faster in case of the phosphate buffer ($\text{pH}=6.8$),
where the protein net charge is around $+7$ \cite{Kuehner1999}. For a certain
amount of the added low-molecular-mass salt, the stability of the
solutions (as indicated by $B_\textrm 2^*$ values) decreases in the
order: $\text{Cl}^- >  \text{Br}^- >  \text{I}^-$. This is the so-called
inverse Hofmeister series, which has been observed
experimentally in several papers \cite{Zhang2009,Boncina2008}.
Unfortunately, the experimental
results for the exact conditions studied in figure~\ref{B2-figs}
are not available so far.

\section{Concluding remarks}

We present new measurements of the cloud-point temperature,
$T_\textrm {cloud}$, for various lysozyme-buffer-salt mixtures. The
salts mixed with the protein were NaSCN, NaI, NaNO$_3$, NaBr,
and NaCl in phosphate buffer ($\text{pH}=6.8$) and NaSCN,
NaI, NaNO$_3$, and KBr in acetate buffer ($\text{pH}=4.6$).
Our measurements, in agreement with some previous studies,
suggest strong salt-specific effects. The $T_\textrm {cloud}$ values,
after a certain amount of buffer is added, do not depend any
more on the total ionic strength of the present electrolyte
($I_\textrm {buffer}+ I_\textrm {salt}$) but rather on its composition; i.e.,
on $I_\textrm {salt}$ content. The cloud-point temperature values can
be modelled as a function of the square root of $I_\textrm {salt}$; cf.
equation (\ref{Correlation-eq}). From the measurements we extracted an
information on the protein-protein interaction under conditions
where $I_\textrm {salt}$ varies. Using this information, we predicted
the relevant liquid-liquid phase diagrams and reduced second
virial coefficients.  The critical temperature of the phase
diagram increases with an increasing salt content, but for
iodide salts, the effect is much stronger than for bromide salts.
This holds true for both buffers studied here. The results for
reduced second virial coefficients, $B_\textrm 2^*$, are consistent
with these observations: the  addition of iodide salt
destabilizes the protein solutions more than the addition of
bromide salt. We believe that the reason lies in a relatively
high (comparing to Cl$^-$ or Br$^-$ ions) hydration free energy
of I$^-$ ion, which is prone to release some hydration water
upon binding to the protein charges. This assumption is
supported by figure~\ref{Correlation-G}, where the correlation
between the strength of the protein-protein interaction and
the free energy of solvation of various counterions is shown.

\newpage

\section*{Acknowledgements}
This study was supported by the Slovenian Research Agency fund
through the Program 0103--0201 and by NIH research grant
GM063592.  Miha Kastelic and Tadeja Janc were supported by the
ARRS Grants for Young researchers.



\clearpage

\ukrainianpart

\title{Ефекти специфіки солі в розчинах лізозиму}
\author{Т. Янч, М. Кастеліч, М. Бончіна, В. Влахі}
\address{Факультет хімії і хімічної технології, Університет Любляни, вул. Вечна, 113, 1000 Любляна, Словенія}

\makeukrtitle

\begin{abstract}
\tolerance=3000%
Ефекти додавання солей з малою молекулярною масою на властивості водних розчинів лізозиму
встановлюються з вимірювань температури точки хмари, $T_\text{cloud}$. Досліджуються суміші протеїну, буферного електроліту і простої солі у воді при $\text{pH} = 6.8$ (фосфатний буфер) і $\text{pH} = 4.6$ (ацетатний буфер). Ми показуємо, що додавання буферного електроліту вище $T_\text{buffer} = 0.6$~моль~дм$^{-3}$ не впливає на
значення $T_\text{cloud}$. Однак, при заміні певної кількості буферного електроліту іншою сіллю, тримаючи постійною повну іонну міцність, ми можемо суттєво змінити температуру точки хмари. Усі солі дестабілізують розчин, і величина ефекту залежить від природи солі.
Експериментальні результати аналізуються в рамках однокомпонент\-ної моделі, яка розглядає
взаємодію протеїн-протеїн як сильно направлену та короткодіючу. Ми використовуємо цей підхід для передбачення других віріальних коефіцієнтів та фазових діаграм рідина-рідина при умовах, коли $T_\text{cloud}$ визначається експериментально.

\keywords лізозим, ефекти специфіки солі, температура точки хмари

\end{abstract}

\end{document}